\documentclass[preprint,aps]{revtex4}

\usepackage{graphicx}
\usepackage{dcolumn}
\usepackage{bm}
\usepackage{natbib}
\usepackage{amsmath}
\usepackage{here}

\begin{document}

\title{InAs Nanowire MOS Capacitors}

\author{Stefano Roddaro}
\affiliation{the Nanometer Structure Consortium, Lund University, P.O. Box 118, 22100 Lund, Sweden}
\author{Kristian Nilsson}
\affiliation{the Nanometer Structure Consortium, Lund University, P.O. Box 118, 22100 Lund, Sweden}
\author{Gvidas Astromskas}
\affiliation{the Nanometer Structure Consortium, Lund University, P.O. Box 118, 22100 Lund, Sweden}
\author{Olov Karlstr\"om}
\affiliation{the Nanometer Structure Consortium, Lund University, P.O. Box 118, 22100 Lund, Sweden}
\affiliation{Mathematical Physics, Lund University, P.O. Box 118, 22100 Lund, Sweden}
\author{Andreas Wacker}
\affiliation{the Nanometer Structure Consortium, Lund University, P.O. Box 118, 22100 Lund, Sweden}
\affiliation{Mathematical Physics, Lund University, P.O. Box 118, 22100 Lund, Sweden}
\author{Lars Samuelson}
\affiliation{the Nanometer Structure Consortium, Lund University, P.O. Box 118, 22100 Lund, Sweden}
\author{Lars-Erik Wernersson}
\affiliation{the Nanometer Structure Consortium, Lund University, P.O. Box 118, 22100 Lund, Sweden}

\date{\today}

\begin{abstract}
We present a capacitance-voltage study for arrays of vertical InAs nanowires. MOS capacitors are obtained by insulating the nanowires with a conformal $10\,{\rm nm}$  ${\rm HfO_2}$ layer and using a top Cr/Au metallization as one of the capacitor's electrodes. The described fabrication and characterization technique enables a systematic investigation of the carrier density in the nanowires as well as of the quality of the MOS interface.   
\end{abstract}
\pacs{to be added}

\maketitle

The development of wrap-gate nanowire (NW) field effect transistors (FETs) is opening promising perspectives for future high-performance electronic devices~\cite{WigFET,GeSiFET}. NWs allow the integration of semiconductor materials with reduced lattice-matching constraints~\cite{Strain1,Strain2} and offer the intriguing possibility of growing III-V structures on Si substrates, thus introducing high-mobility and optically-active elements on a Si platform~\cite{NWonSi}. However, many of the key parameters of the NWs such as doping level and carrier distribution are still difficult to determine in a direct and conclusive way. For conventional FETs it is possible to take advantage of capacitance-voltage (CV) characterizations to determine, in a precise way, carrier concentration and interface properties of planar metal-oxide-semiconductor (MOS) stacks. Similar measurements have been largely unavailable for semiconductor NWs because of the extremely small capacitance of these nanostructures (down to aF). Recent experimental studies showed that such a small capacitance can be detected using bridge measurements together with appropriate screening~\cite{SingleCV}. Here we demonstrate CV measurements of small arrays of vertical NWs, where the NW capacitance can be easily separated from the parasitic capacitance between the gate connection and the conducting substrate. Our vertical fabrication protocol is scalable and thus enables parallel processing, which is crucial for a systematic investigation of the device properties.

\begin{figure}[ht!]
\begin{center}
\vspace{0.3cm}
\includegraphics[width=0.46\textwidth]{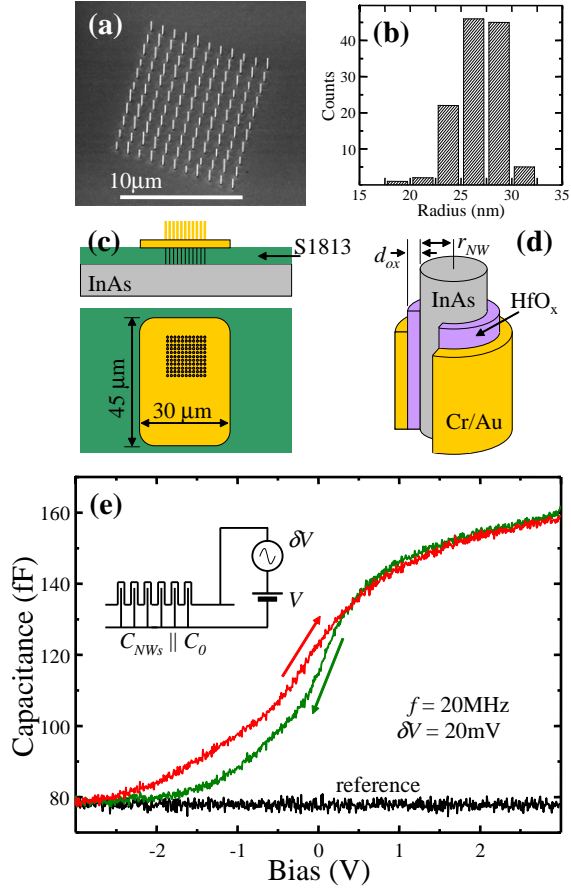}
\caption{(a) Scanning electron micrograph of a $11\times11$ InAs nanowire array (tilt angle $52^\circ$). (b) Typical radius distribution in the array. (c) and (d) Details of the device structure. (e) Representative $C(V)$ scan from $-3\,{\rm V}$ to $+3\,{\rm V}$ (red) and return (green) compared with a bare pad scan (black).}
\end{center}
\end{figure}

The device structure is presented in Fig.~1. NW arrays (Fig.~1a) were obtained by self-assembled growth in a chemical beam epitaxy (CBE) system. NW fomartion is guided by gold nanoparticles that are deposited on a doped InAs (111)B substrate~\cite{CBE}. A number of arrays was defined in parallel with various nanoparticle sizes to study radius dependance. For the present investigation $5$ different groups of $15$ nominally identical NW arrays were fabricated with an average radius $r_{N\!W}$ of $23.0\,{\rm nm}$, $25.0\,{\rm nm}$, $26.5\,{\rm nm}$, $28.5\,{\rm nm}$ and $30.0\,{\rm nm}$, respectively. Panel (b) shows a typical radius distribution in a single $11\times11$ array with a standard deviation of about $4.0\,{\rm nm}$. The device structure is sketched in panel (c) and (d): NWs were first insulated by a conformal ${\rm HfO_2}$ coating (purple) by atomic layer deposition (125 cycles at $250\,^\circ{\rm C}$, corresponding to $d_{ox}\approx10\,{\rm nm}$); the top electrode encapsulating the NWs was then fabricated by sputtering a Cr/Au bilayer (nominal $20/25\,{\rm nm}$). A polymeric film of S1813 from Shipley with a thickness of about $1\,{\rm \mu m}$ (green) was used as a lifting layer in order to increase the ratio $C_{N\!W}/C_0$ between the NW capacitance $C_{N\!W}$ and stray capacitance $C_0$ in our devices. Single devices were finally defined by UV lithography and metal etching of $30\times45\,{\rm \mu m^2}$ gate pads.
 
The NW capacitance was measured at room temperature in a Cascade probe station system equipped with an Agilent 4294A impedance analyzer. The complex impedance $Z=|Z|e^{i\theta}$ was measured using a small AC modulation $\delta V = 20\,{\rm mV}$ on top of a DC bias $V$ in the range $\left[-3\,{\rm V},+3\,{\rm V}\right]$. A simplified scheme of the biasing configuration is shown in the inset to Fig.~1(e). The measured $Z$ was found to be mostly capacitive ($\theta\approx-90^\circ$) and was interpreted in terms of a series $RC$ model with $Z = R-i/\omega C$. Such a simple model is appropriate in our case and experimental $Z(\omega,V)$ data for $V\gtrsim+1\,{\rm V}$ yield a frequency-independent and well-defined $C(V)$. The frequency evolution of $Z(\omega,V)$ in the depletion regime for $V<0$ is less trivial as expected due to the increasing NW resistance, to the activation of slow trap states at the interfaces and to effects of inversion in the InAs semiconductor. In particular, the increasing importance of $RC$ constants close to the pinch-off is a peculiarity of our cylindrical geometry and sets a qualitative difference with respect to conventional planar MOS capacitors. The plot in Fig.~1(e) shows typical $C(V)$ sweeps obtained on devices from the group $r_{N\!W}=26.5\,{\rm nm}$ at a frequency $f=20\,{\rm MHz}$: we mark the sweep going from negative to positive $V$ as $C_\uparrow(V)$ (red) and $C_\downarrow(V)$ for the opposite sweep direction (green). The capacitance saturates at negative voltages to $C_0\approx70-80\,{\rm fF}$, grows sharply across $V\approx0\,{\rm V}$ and flattens again for $V>1\,{\rm V}$ in the accumulation regime. Differently from conventional MOS capacitors, here we expect the NW to become insulating in the depletion limit and $C$ to approach {\em zero} instead of a finite depletion capacitance. Indeed, here the observed saturation $C\to C_0$ corresponds to the NW depletion, as proved by comparison with four bare pads of the same geometry (black curve). The presence of $C_0$ is not linked to the NWs and it is rather due to both the parallel capacitance between the pad and the substrate as well as the one between the probe tips and the substrate.

Hysteresis effects are analyzed in Fig.~2. In the first panel, the shift between the capacitances measured in the two opposite sweep directions is barely visible on small (less than $1\,{\rm V}$) sweep ranges while it increases for larger $V$ swings. $C_\downarrow(V)$ curves do not depend strongly on the DC sweep swing while $C_\uparrow(V)$ curves tends to move towards higher $C$ values (or lower $V$ values, for a given $C$) when the sweep is extended from $\pm0.5\,{\rm V}$ up to $\pm3.0\,{\rm V}$. The shift between $C_\uparrow$ and $C_\downarrow$ does not depend strongly on the sweep speed (about $150\,{\rm mV/s}$ in our case) and time-dependent measurements indicate that capacitances tend to relax from $C_\uparrow(V)$ towards $C_\downarrow(V)$ on a timescale $\tau\approx 30\,{\rm mins}$. We conclude that $C_\downarrow(V)$ results from an equilibrium distribution of charges at the capacitor's interfaces while a long-lived out-of-equilibrium distribution is present along $C_\uparrow(V)$. This effect can be evaluated quantitatively in a simple way if one assumes that trapped charges are located exactly at the NW surface: in that case the addition of a surface charge density $\Delta\sigma_s$ will shift an ideal $C(V)$ curve as

\begin{figure}[ht!]
\begin{center}
\includegraphics[width=0.46\textwidth]{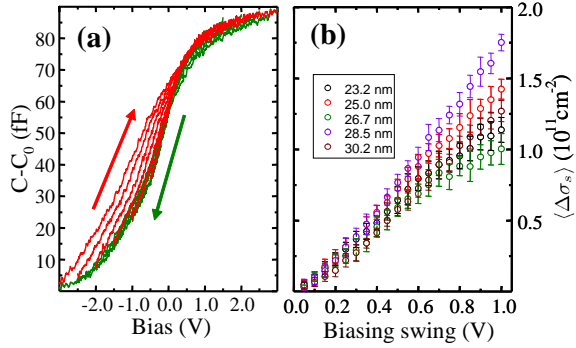}
\caption{(a) Evolution of the hysteresis cycle for increasing gate swings from $[-0.5,+0.5]\,{\rm V}$ up to $[-3.0V,+3.0]\,{\rm V}$. (b) Charge loop integrals $\langle\Delta\sigma_s\rangle$ for different device geometries as a function of the biasing swing around $0\,{\rm V}$.}
\end{center}
\end{figure}

\begin{equation}
C_{meas}(V)=C(V+ S_{N\!W}\times\Delta\sigma_s/C_{ox})
\end{equation}

\noindent where $S_{N\!W}=2\pi r_{N\!W}L_{N\!W}$ and $L_{N\!W}$ are the surface and length of the gated NW, respectively, while $C_{ox}$ is the oxide capacity $2\pi\varepsilon L_{N\!W}/\log(1+d_{ox}/r_{N\!W})$. The value of $\Delta\sigma_s$ depends on the biasing history of the device, thus we obtain the different hysteresis cycles for different sweep swings. Figure~2b shows the average surface charge 

\begin{equation}
\langle\Delta\sigma_s\rangle=\frac{C_{ox}}{S_{N\!W}}\times\frac{1}{\Delta C}\oint CdV,
\end{equation}

\noindent where $\Delta C$ is the capacitance swing of the cycle and we used an average $L_{N\!W}=680\,{\rm nm}$ (from SEM imaging of the devices), $\varepsilon=15\varepsilon_0$~\cite{LEInAsCap,HfORef} and $C_{ox}=1.78\,{\rm fF}$. The plot reports the loop integrals for the various device groups we studied: for $V$ swings below $\approx 0.5\,{\rm V}$ we obtain $\langle\Delta\sigma_s\rangle<1.0\times 10^{11}\,{\rm cm^{-2}}$, which seems very promising for device applications of NW as wrap-gate transistors~\cite{missing}. Note however, that the hysteresis in surface charge becomes much
larger if the bias sweep extends further into the depletion region.

\begin{figure}[ht!]
\begin{center}
\includegraphics[width=0.46\textwidth]{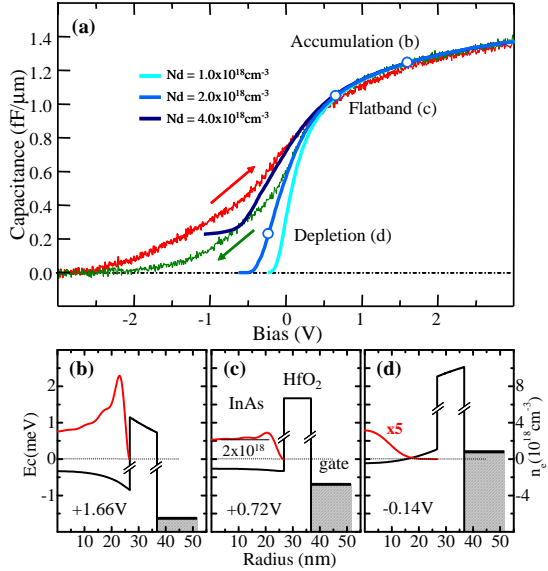}
\caption{(a) Theoretical fit of the dataset $C_\downarrow(V)$ of Fig.~1(e) using three different carrier densities. The band bending (black) and electron density (red) at the three different points along the blue line at doping $N_d=2\times10^{18}\,{\rm cm^{-3}}$ are reported in the lower panels for accumulation (b), flatband (c) and depletion (d) conditions.}
\end{center}
\end{figure}

To further analyze the data, we performed detailed calculations for the capacitance on the basis of a Poisson-Schr\"odinger code similar to Refs.~\cite{Theo,Theo2}. Fig.~3(a) shows the unit length capacitance for three different doping densities $N_d$ of the wire, which are treated as a homogeneous positive background charge. The experimental data shown correspond to the assumption that $90$ out of $121$ wires are actually properly connected in the device: this scaling is required in order to match the geometry-set capacitance in accumulation and is not unreasonable given the present device parameters. The best fit is obtained using a doping of $2.0\times10^{18}\,{\rm cm^{-3}}$: the curve at $1.0\times10^{18}\,{\rm cm^{-3}}$ rolls down too quickly with the voltage $V$; differently at $N_d=4.0\times10^{18}\,{\rm cm^{-3}}$ the valence bands cross the Fermi level at the interface before the conduction band is completely depleted ($0.54\,{\rm eV}$ was used as the wurtzite InAs gap~\cite{HolesW}) and screening effects due to inversion are expected to show up, inconsistently with observations. It is interesting to note that all the fit curves in Fig.~3a fall nearly $50\%$ short of the classical $C_{ox}=2.61\,{\rm fF}$ (for a $1.0\,{\rm \mu m}$ length, $r_{N\!W}=26.5\,{\rm nm}$ and $d_{ox}=10\,{\rm nm}$) even in the accumulation regime at $V\approx+3.0,{\rm V}$. This is an effect of quantum capacitance is due to the narrow radius of the NW with respect to the screening length. Lower panels indicate the corresponding conduction band diagram $E_c(r)$ in the capacitor in the accumulation (b), flatband (c) and depletion (d) regimes: the corresponding positions along the $C(V)$ fit are indicated in the top panel. We obtained best agreement assuming the gate bias $V$ to be $0.39\,{\rm V}$ larger than the calculated electrostatic potential at the gate. This shift can be attributed to the difference
between the work function of Cr ($4.5\,{\rm eV}$) and the electron affinity of InAs ($4.9\,{\rm eV}$ for zincblende lattice) as well as negative
fixed charges (with areal density $\sim 8\times 10^{12}\,{\rm cm^{-2}}$) trapped in the oxide. As the electron affinity of the nanowire is uncertain due to the uncommon wurtzite structure exhibiting a larger band gap~\cite{HolesW}, this estimate for the density of fixed charges is probably too large. It is crucial to note here that we assumed that only electrons in the conduction band are able to contribute to the $C(V)$ at our frequency. We interpret the discrepancy between fit and experiments for $V<0\,{\rm V}$ as due to the effect of screening of slow trap states in the InAs gap~\cite{TrapInTheGap}, which indeed start becoming important at $V\approx0\,{\rm V}$ in our simulations. Consistently with this interpretation, we observed experimentally that such discrepancies becomes larger as the frequency is decreased and a clear $C(V)$ step develops, similarly to what has been reported in previous studies on planar structures~\cite{Step}. 

In conclusion we have demonstrated a technique for capacitance-voltage characterizations of arrays of vertical InAs NWs. Our analysis allows evaluating the role of surface states as well as yields an estimate of the doping in the NW, thanks to a detailed comparison with Poisson-Schr\"odinger simulations. Preliminary results indicate promising device parameters in view of the application of wrap-gate NWs as high-performance transistors.

This work was supported by the Swedish Research Council, the Swedish Foundation for Strategic Research, the EU-project NODE 015783, the Knut and Alice Wallenberg Foundation and the Italian Ministery of University and Research.

\end{document}